\documentclass[conference,a4paper]{APSIPA2021}
\usepackage{amsmath}
\usepackage{graphicx}
\usepackage{multirow}
\usepackage{threeparttable}
\usepackage[backend=biber,style=ieee,]{biblatex}
\usepackage{geometry}
\usepackage{algorithm}
\usepackage{algorithmicx}
\usepackage{algpseudocode}  
\usepackage{textcomp}
\usepackage{xcolor}
\usepackage{utfsym}
\usepackage{fancyhdr}

\fancypagestyle{firststyle}{
  \fancyhf{}
  \fancyhead[C]{2023 Asia Pacific Signal and Information Processing Association Annual Summit and Conference (APSIPA ASC)}
}

\begin{document}

\title{NADiffuSE: Noise-aware Diffusion-based Model for Speech Enhancement}

\author{
\authorblockN{
Wen Wang, 
Dongchao Yang, 
Qichen Ye, 
Bowen Cao and 
Yuexian Zou$^{\ast}$\thanks{*Corresponding author}
}


\authorblockA{
the School of Electronic and Computer Engineering, Peking University, Shenzhen, 518055, China \\
E-mail: \{wangw, 2001212832, yeeeqichen, cbw2021\}@stu.pku.edu.cn, zouyx@pku.edu.cn}
}

\maketitle
\renewcommand{\thefootnote}{\fnsymbol{footnote}} 
\footnotetext[1]{Corresponding author.} 

\pagestyle{empty}
\thispagestyle{empty}

\begin{abstract}
  The goal of speech enhancement\,(SE) is to eliminate the background interference from the noisy speech signal. Generative models such as diffusion models\,(DM) have been applied to the task of SE because of better generalization in unseen noisy scenes. Technical routes for the DM-based SE methods can be summarized into three types: task-adapted diffusion process formulation, generator-plus-conditioner\,(GPC) structures and the multi-stage frameworks. We focus on the first two approaches, which are constructed under the GPC architecture and use the task-adapted diffusion process to better deal with the real noise. However, the performance of these SE models is limited by the following issues: (a) Non-Gaussian noise estimation in the task-adapted diffusion process. (b) Conditional domain bias caused by the weak conditioner design in the GPC structure. (c) Large amount of residual noise caused by unreasonable interpolation operations during inference. To solve the above problems, we propose a noise-aware diffusion-based SE model\,(NADiffuSE) to boost the SE performance, where the noise representation is extracted from the noisy speech signal and introduced as a global conditional information for estimating the non-Gaussian components. Furthermore, the anchor-based inference algorithm is employed to achieve a compromise between the speech distortion and noise residual. In order to mitigate the performance degradation caused by the conditional domain bias in the GPC framework, we investigate three model variants, all of which can be viewed as multi-stage SE based on the preprocessing networks for Mel spectrograms. Experimental results show that NADiffuSE outperforms other DM-based SE models under the GPC infrastructure. Audio samples are available at: https://square-of-w.github.io/NADiffuSE-demo/.
\end{abstract}

\section{Introduction}
\label{intro}
In the last decade, deep learning\,(DL) methods \cite{xu2013SE_overview,overview2} have become the mainstream for speech enhancement\,(SE), which can be divided into two categories: the discriminative and generative ones \cite{zhou2021machine}. The discriminative methods learn nonlinear mappings\cite{mapping} or estimate time-frequency masks \cite{ibm,irm,cirm} through annotated speech data pairs, but suffer from non-linear artifacts and poor generalizations \cite{SE-artifacts}. The generative approaches \cite{segan,sevae,seflow} employ different infrastructures such as Generative Adversarial Networks (GANs) \cite{gan}, Variational Autoencoders (VAEs) \cite{vae} and flow-based models \cite{flow} to learn the distribution of clean speech signals, which can be more robust to complex and varying noise scenarios \cite{GM-goodness}. 

Diffusion Denoising Probabilistic Models\,(DDPM) \cite{DDPM} is a new kind of generative model inspired by the nonequilibrium thermodynamics \cite{DDPM-inspiration}. DDPM corrupts the data to the pre-defined Gaussian distribution by gradually adding noise at the forward process, and the random noise is progressively denoised and finally restored to the original data at the reverse stage. Many works \cite{kong2020diffwave,chen2020wavegrad,diffsound} have employed the DDPM to generate natural and high-quality human voice from the given text or Mel spectrogram. Recently, DDPM-based SE methods \cite{SGMSE,sgmse+,cdiffuse,universal,modified,diffuse,srtnet,StoRM,SERefiner} have also been explored for the promising result in speech generation.

\begin{figure}
\centering
\centerline{\includegraphics[width=\columnwidth]{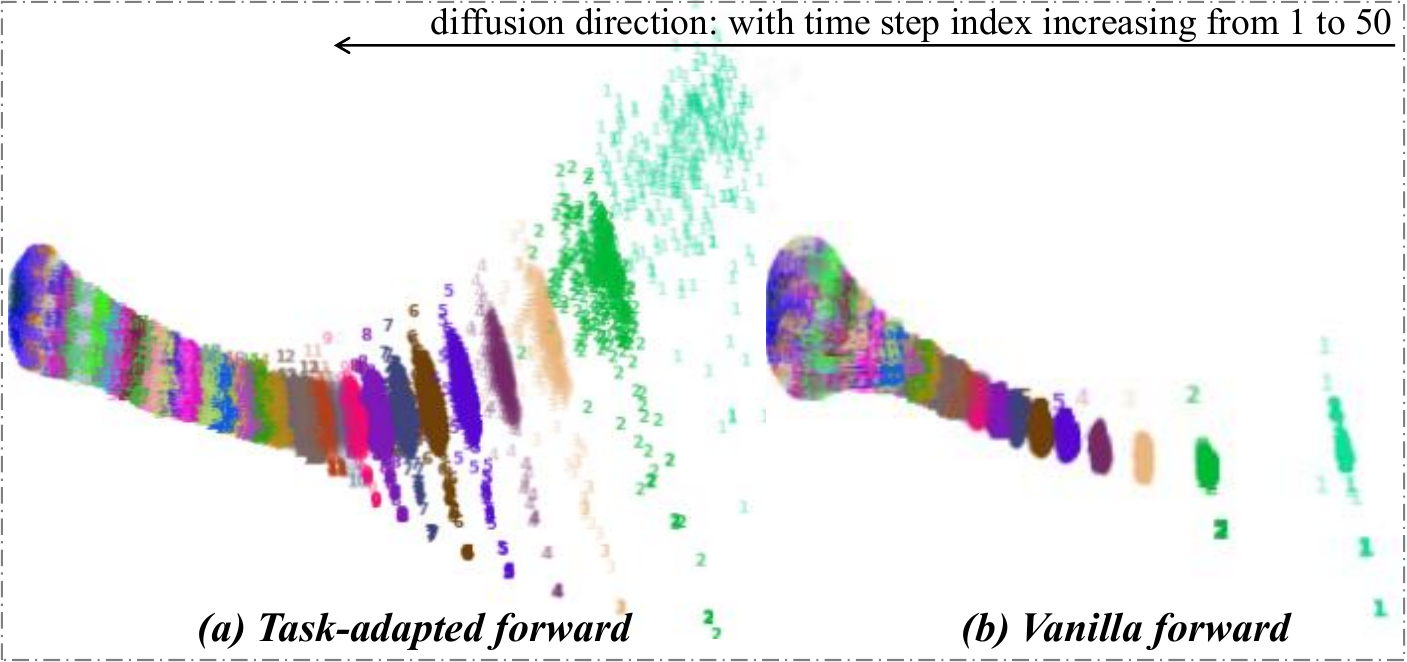}}
\caption{To demonstrate the effect of non-Gaussian noise in diffusion-based SE, we visualize the vanilla and task-adapted diffusion process: we randomly selected 2000 utterances, which are first converted into Mel spectrograms, then cropped to 62 frames (same as the training settings), and finally visualized in two dimensions using the t-SNE algorithm. We use different colors and indexes to label the data at different time steps.}
\label{forward} 
\end{figure}
Current diffusion-based SE models can be categorized into three types.
The first \cite{cdiffuse,SGMSE,sgmse+} is to change the mathematical form of the diffusion process to adapt to the task of SE, in which the mean of clean speech signals is gradually pulled towards the noisy ones by interpolating the asymptotically increasing noisy signal, so that the reverse stage is directly the speech enhancement process. The second \cite{universal,modified,diffuse,cdiffuse} is to train a conditional network based on a well-trained pure speech generation network\,(Generator-plus-Conditioner, GPC). Under this setting, the enhanced speech signal is generated with the acoustic feature produced by the conditioner, which can be expressed as \textbf{\textit{output = generator(conditioner(noisy input))}}. The last is to develop a multi-stage SE\, where the diffusion model aims to learn the fine-grained or residual speech signal based on the coarsely enhanced one \cite{srtnet,StoRM,SERefiner}. This study focuses on the first two lines, and leverages a task-adapted diffusion process under the GPC architecture to deal with the real noise.

However, there exists three problems in the existing works: (1) \textbf{Non-Gaussian Estimation}. As shown in Fig. \ref{forward}, incorporating the noisy signal into the forward process makes the data no longer satisfies the tight Gaussian distribution at each step, which we attribute to the effect of background noise. Noise interference varies in realistic scenarios, hence it is challenging for SE models to understand the patterns of various noise and to adaptively learn the ability of denoising in real situations \cite{noise-aware1,noise-aware2,noise-aware3,noise-aware4}. Moreover, the estimation target changes from the added Gaussian noise to a combination of Gaussian noise and non-Gaussian background noise, which makes training the model harder \cite{SGMSE}. (2) \textbf{Conditional domain bias}. In the GPC framework, the generator and conditioner determine the upper and lower bounds for the SE performance: If a conditioner can ideally map the noisy features completely to the pure ones, the best performance of SE is then obtained\,(upper bound). Otherwise, if the conditioner does not work at all, the worst performance is obtained\,(lower bound). Table \ref{t1} records the SE performance bounds of \cite{cdiffuse}. We observe that current conditioner is not sufficient to compensate for the gap between lower and upper bounds. We define it as conditional domain bias, which results from the change in the dimension, type and purity of the acoustic features (usually Mel spectrogram). (3) \textbf{Under-explained interpolation}. In the original inference algorithm\,(\emph{cf.} Line 11, Alg. \ref{alg:inf}), the linear interpolation operation \cite{diffuse,cdiffuse,SGMSE} is commonly used in the last step with a certain percentage ($r=0.2$), which serves as an implicit post-processing method to supplement the lost speech details. 
Although the interpolation does improve the objective evaluation metrics (\emph{cf.} Table \ref{tab:inf} row 4 and 5), its validity is limited due to the introduced large amount of additional noise.
We give a comparison between with and without the interpolation in Fig. \ref{melset} as an example of evidence: the white dashed box in (f) indicates the presence of the significant noise components.

To address the above issues, our improvements are refined into a noise-aware diffusion-based SE model\,(NADiffuSE). The main contributions of this work are summarized as follows:
\begin{itemize}
    \item[1.] To more accurately estimate the non-Gaussian noise component in the task-adapted diffusion process, we propose to use noise encodings to guide the diffusion model for adaptive noise reduction in real noisy situations.
    \item[2.] To further reduce the conditional domain bias under the GPC architecture, we design three network variants based on the additional pre-processor network to improve the quality of regenerated speech signals.
    \item[3.] To reduce the additional noise introduced by the interpolation operation, we construct a relatively accurate data anchor point from noisy speech signals at specified time steps, based on which we use iterative interpolation operations to refine the speech details.
\end{itemize}

\section{Related Work}
\subsection{Diffusion Model}
The diffusion model \cite{DDPM,ddim,SGM} contains a forward and a backward process. In the mathematical form of DDPM \cite{DDPM}, each process is represented by a first-order Markov chain with fixed time steps. In the forward process, a series of corrupted data $x_{1:T}=x_1,x_2,...,x_T$ can be obtained from the original clean data $x_0$ through the transfer distribution $q(x_t|x_0) = \sqrt{\Bar{a}_t}x_0+\sqrt{1-\Bar{a}_t}\epsilon$, where $\Bar{a}_t=\prod\limits_{i=1}^T a_i$, $a_t = 1-\beta_t$, $\epsilon\sim\mathcal{N}(0,I)$ and $\beta_t$ is a constant defined in advance. In the reverse process, the original data $x_0$ can be recovered from the latent distribution $x_T$ by iteratively performing the backward transfer step $ p_\theta(x_{t-1}|x_\text{t})=\mu_\theta(x_t)+\widetilde{\beta_t}I$, where $\widetilde{\beta_t}$ is a constant and $\mu_\theta(x_t) = \frac{1}{\sqrt{a_t}}(x_t-(\beta_t/\sqrt{1-\Bar{a}_t})\epsilon_\theta(x_t))$.

\subsection{Diffusion-based SE Methods}
The current technical routes of diffusion-based SE mothods can be divided into three categories: Task-adapted diffusion formula, GPC structures and multi-stage frameworks.

\textbf{Task-adapted diffusion formula} \cite{cdiffuse,SGMSE,sgmse+} adapt the task of SE by incorporating the noisy sigal $y$ into the diffusion process: the mean of $x_t$ is progressively pulled from the $x_0$ to $y$. The data degradation in \cite{cdiffuse} is formulated as Eq. \ref{cforward}, where $m_t=\sqrt{(1-\Bar{\alpha_t})/\sqrt{\Bar{\alpha_t}}}$ denotes the asymptotic coefficient and $\Bar{\delta_t}=1-(1+m_t^2)\Bar{\alpha_t}$ is the variance for added Gaussian noise.
\begin{equation}\label{cforward}
x_t = \sqrt{\Bar{a_t}}((1-m_t)x_0+m_t\cdot y)+\sqrt{\Bar{\delta_t}}\epsilon
\end{equation}
While \cite{SGMSE,sgmse+} have the closed-form solution for the mean value as $\mu (x_0,y,t) = e^{-\gamma t}x_0+(1-e^{-\gamma t})y$.

\begin{table}[t]
\begin{center}
    \caption{ Evaluation results of the GPC structure \cite{diffuse} on new\,(a) and original\,(b) VoiceBank-Demand; For WaveNet\cite{oord2016wavenet}-based generator we evaluated the SE performance's Upper\,(generator with clean Mel spectrograms) and Lower\,(with noisy Mel spectrograms) bounds. Task-adapted diffusion process \cite{cdiffuse} is adopted here.}
    \label{t1}
    \begin{tabular}{cccccc}
    
    \multicolumn{6}{c}{(a) Evaluation results on new VoiceBank-Demand}\\
    \hline
     Model & Mode & CSIG & CBAK & COVL & PESQ \\
    \hline
    Unprocessed & / &3.65&3.16&2.91&2.13\\
    CDiffuSE &/  &3.83&3.13&3.19&2.55\\
     \hline
     \multirow{3}*{DiffWave}& Upper &4.41&3.57&3.84&3.26\\
     & Lower &3.45&2.68&2.78&2.16\\
     \hline
     \multirow{3}*{DiffWave-Cls}& Upper &\textbf{4.65}&\textbf{3.67}&\textbf{4.05}&\textbf{3.44}\\
     & Lower &3.55&2.73&2.86&2.20\\
     \hline
     \multirow{3}*{DiffWave-Emb}& Upper &4.34&3.54&3.79&3.22\\
     & Lower &\textbf{3.59}&\textbf{2.78}&\textbf{2.89}&\textbf{2.22}\\
     \hline
     \\

     \multicolumn{6}{c}{(b) Evaluation results on original VoiceBank-Demand}\\
     \hline
     Model & Mode & CSIG & CBAK & COVL & PESQ \\
     \hline
     Unprocessed & / &3.35&2.44&2.63&1.97\\
     CDiffuSE &/ &3.66&2.83&3.03&2.44\\
     \hline
      \multirow{3}*{DiffWave}& Upper &4.34&\textbf{3.53}&\textbf{3.79}&\textbf{3.21}\\
     & Lower &3.40&\textbf{2.55}&2.71&2.08\\
     \hline
      \multirow{3}*{DiffWave-Emb}& Upper &\textbf{4.36}&3.42&3.78&3.19\\
     & Lower &\textbf{3.41}&2.47&\textbf{2.74}&\textbf{2.15}\\
     \hline
     \end{tabular}    
\end{center}
\end{table}


\begin{figure*}[t]
\label{overview}
\centering
\centerline{\includegraphics[scale=0.65]{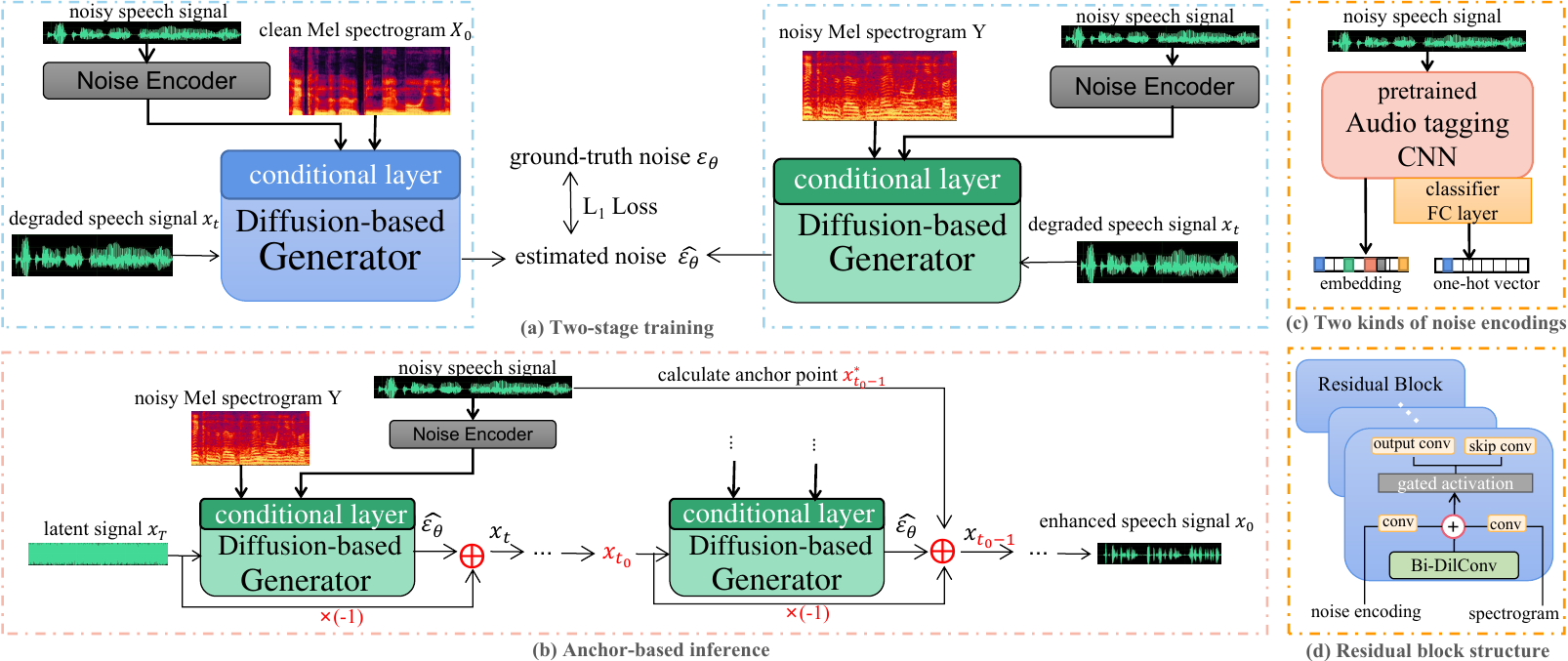}}
\caption{An overview of the proposed NADiffuSE: (a) pre-training (left) and fine-tuning (right), both of which are conditioned on Mel spectrograms and noise encodings; (b) improved inference process where the iterative interpolation operates in the last $t_0$ steps based on the anchor point; (c) latent embedding (left) and category classifier (right) of noise encoding in proposed noise-aware training; (d) bi-conditional Residual blocks.}
\label{fig:overview}
\end{figure*}
\textbf{GPC structures} consist of a generator which can generate the clean speech signal based on the clean conditions and a conditioner, i.e. network designs or training methods. One kind of conditioner \cite{diffuse,cdiffuse} works through two-stage training: the first stage uses the 80-dim pure Mel spectrogram as a condition\,(pre-training), and the second one adjusts the weights using the 513-dim noisy amplitude spectrogram\,(fine-tuning). The other \cite{modified} is replacing the submodule in the generator with a newly trained one to alter the degraded conditions to match the original ones.

\textbf{Multi-stage frameworks} take full advantage of the nature that the diffusion model is more suitable for detail refinement. The coarsely enhanced result is obtained through a discriminative model or also a diffusion-based mothod. And then conditioned on the coarsely enhanced signal, the diffusion model is used to generate the fine-grained enhanced signal \cite{SERefiner,StoRM} or the residual signal \cite{srtnet} against the clean one.

\section{Proposed Method}
We make improvements to address the above issues and give an overview diagram in Fig. \ref{fig:overview}. The training process in (a) follows the same two-stage paradigm as \cite{cdiffuse}, where both noise encodings and spectrograms are used as conditional information. The inference process in (b) is iterative and improved using noisy speech signals $y$ in the last steps.

\subsection{Noise encodings}
We propose to additionally use noise coding as global conditional information in an aim to mine the priori information of the acoustic noise to guide the diffusion model for accurate combine noise estimation. We explore two different types of noise encodings as drawn in (c). 

\subsubsection{Category classifier}
\label{method:classifier}
We first use the categorical properties of the background noise, in which the noise type labels are fed into the learnable embedding layer in the form of one-hot vectors. In order to get the ground truth noise label during training, we only consider a fixed number of closed sets of noise scenes. At the inference stage, a noise classifier is required to predict the noise type of the noisy speech signal. The background noise is usually labeled using the the keywords of the acoustic scene where the noise is collected, like living room, street, bathroom and etc. The concept of noise semantics is difficult to define and we propose to describe the background noise with acoustic events. We observe a one-to-many relationship between noise scenes and sound events, for example, keyboard tapping, TV background and babble noise may exist simultaneously in a living scene. Therefore, we perform transfer learning based on a large-scale pre-trained network structure \cite{panns} for the audio tagging task. We load the pre-trained weights as initialization and add a linear classification into the original structure. As described above, we build a noise classifier whose input is the noisy speech signal and output is a value belonging to a predefined set of noisy scene labels $\left\{0,1,...,N-1\right\}$, with $N$ being the total number of noise categories.

\subsubsection{Latent embedding}
In order to extend the noise encodings to the open domain, we further explore to characterize the noise properties in a latent embedding mechanism. We directly use the output of the classifier's (described above) last hidden layer as a noise feature. The 2048-dimensional embedding is extracted from the noisy speech signal and aligned with the hidden dimension of the residual block through MLP structure. It is worth noting that we still fine-tune the convolutional feature extractor using the classification task in order to make it more relevant to the background noise characterization.


\begin{algorithm}[t]
    \caption{Anchor-based Inference algorithm}
    \label{alg:inf}
    \begin{algorithmic}[1]
        \State Sample $x_{T} \sim p_{latent}$
        \For{$t=T-1,T-2,...,0$}
            \State Compute $\mu_{\theta}(x_{t+1},y)$ and $\Bar{\delta_{t}}$ as in Eq. (\ref{reverse})
            \State Sample $x_{t}=\mu_\theta(x_{t+1},y)+\Bar{\delta_{t}}I$ from $p_{\theta}(x_{t}|x_{t+1},y)$
            \If{select Improved Sampling and $t<t_0$:} 
                \State Calculate the anchor point $x_t^*$ using Eq. (\ref{anchor})
                \State Do interpolation $x_t=r_t\cdot x_t^* + (1-r_t)\cdot x_t$
            \EndIf
        \EndFor
        \If{select Original Sampling:} 
            \State $x_0 = r\cdot x_0+(1-r)\cdot y$
        \EndIf
        \State return $x_0$    
    \end{algorithmic}    
\end{algorithm}

\subsection{Anchor-based inference algorithm}
\label{method:inf}
The ideal case of inference is that the reverse data distribution fits the forward one exactly. We use the same conditional diffusion process as in Eq. \ref{cforward}, where the mean value is determined partly by $x_0$ and partly by $y$. Thus we can use the noisy signal $y$ (known during inference) to construct a relatively accurate anchor point for the reverse process in Eq \ref{anchor}, where $\Bar{\alpha_t}$ and $\Bar{\delta_t}$ follow the previous definition.
\begin{equation}\label{anchor}
x_t^*=m_t\sqrt{\Bar{a_t}}y+\sqrt{\Bar{\delta_t}}\epsilon
\end{equation}
Anchors contain both useful clean speech details and degraded background noise. We have described in Section \ref{intro} the pros and cons of the current interpolation operation used in the inference process. Rather than directly interpolating the result using noisy speech at the final step, We use anchors for the interpolation to decrease the extra noise. We use the anchor-based interpolation repeatedly in and only in the last few steps, because the data will be more sensitive to the slight inaccuracies as inference finishes. Therefore, we have used iterative interpolation operations to remove some of the noise residules contained in the anchor points by stepwise refinement. To further weaken the effect of noise, we linearly anneal the interpolation coefficients at the rate of $1/t_0$ (other annealing options are worth exploring). Our improved anchor-based inference algorithm is given in Alg. \ref{alg:inf}: we follow Eq. (\ref{reverse}) at each step to sample $x_{t-1}$ from the reverse probability distribution, and in the last $t_0$ steps perform an interpolation with the anchor point. The mean is a linear combination of $x_t$, $y$ and the estimated noise term $\epsilon$, where $c_{xt}$,$c_{yt}$ and $c_{\epsilon t}$ are constants and the detailed derivation follows \cite{cdiffuse}.
\begin{equation}\label{reverse}
\begin{split}
\mu_\theta(x_{t+1},y)=& c_{xt}x_t+x_{yt}y_t+c_{\epsilon t}\epsilon_\theta\\
\Bar{\delta_t}=& \delta_{t-1} - \frac{1-m_t}{1-m_{t-1}}^2 \alpha_t\frac{\delta_{t-1}^2}{\delta_t}
\end{split}
\end{equation}


\subsection{Model variants in the conditioner design}
\begin{figure}
\centering
\centerline{\includegraphics[width=\columnwidth]{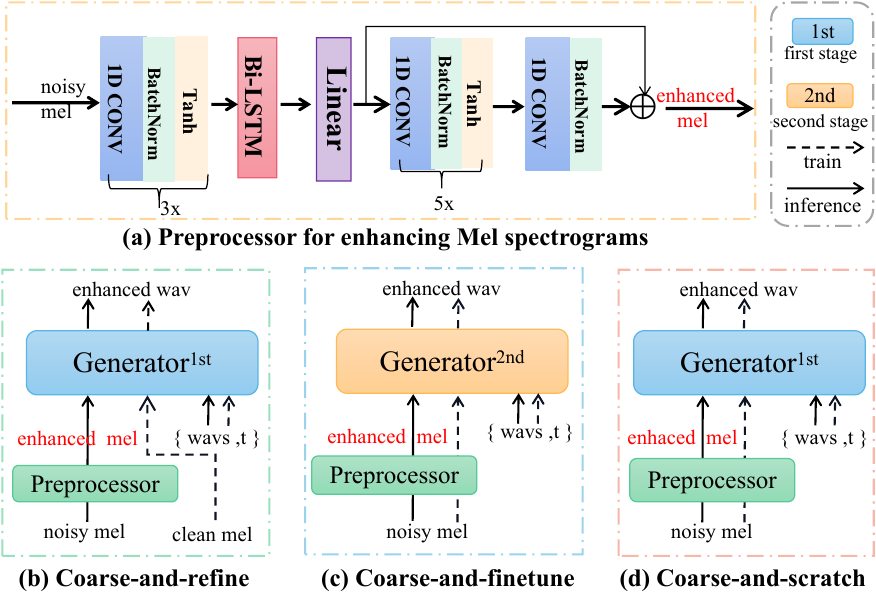}}
\caption{Training\,(dashed line) and inference \,(solid line) pipelines of three conditioners based on the preprocessor: (a) preprocessor used for enhancing the noisy Mel spectrograms\,(known as \textit{coarse}) (b) \textit{coarse-and-refine} only needs one training stage in which the generator is trained with clean Mel spectrograms and inferenced with enhanced ones (c) \textit{coarse-and-finetune} stills goes through the second training stage where the generator is finetuned with enhanced Mel spectrograms (d) \textit{coarse-and-scratch} also needs only one training stage where the generator uses enhanced Mel spectrograms in both the training and inference phases.}
\label{conditioner} 
\end{figure}
In the section \ref{intro} we give a definition of the conditional domain bias, a problem that exists under the generator-plus-conditioner\,(GPC) structure. Following the similar conditioner design in \cite{cdiffuse}, we first consider using the 80-dimensional Mel spectrogram as conditional information in both training stages to reduce the difficulty of aligning the conditional domain in the fine-tuning phase. Given that the current conditional mapping layer contains only a simple up-sampling layer and a shallow MLP structure, we are inspired by \cite{hifangan2} to additionally train a preprocessing network (preprocessor) for enhancing the Mel spectrogram. Thus we can get the pre-enhanced Mel spectrogram and name this process as \textit{coarse}. Given the pre-enhanced Mel spectrogram: on the one hand it can be directly used as the initial result of the enhancement and then refined using the generator\,(coarse-and-refine). On the other, compared to the noisy speech signal's Mel spectrogram, it is less difficult to perform fine-tuning with the pre-enhanced one\,(coarse-and-finetune). Last, the pre-enhanced Mel spectrogram is available in both training and testing phases, which can be used to directly train a generator from scratch, without further fine-tuning. In conclusion, three feasible conditioner designs are illustrated in Fig. \ref{conditioner} (b), (c) and (d). 
\begin{itemize}
    \item[1.] Coarse-and-refine: The generator which is only trained with clean Mel spectrograms can be directly used to generate the speech with the pre-enhanced Mel spectrograms.
    \item[2.] Coarse-and-finetune: After training with clean Mel spectrograms, the generator is then fine-tuned with enhanced Mel spectrograms instead of the noisy ones.
    \item[3.] Coarse-and-scratch: The generator is trained from scratch using the pre-enhanced Mel spectrograms.
\end{itemize}
All three network variants use the pre-enhanced Mel spectrogram as a condition for inference, and Gaussian-like speech signal sampled from latent distribution $N(x_T;\sqrt{\Bar{\alpha}_T}y,\delta_T I)$ as signal input. The above proposed conditioner mechanism is applicable for all SE models under the GPC structure.

\begin{table*}[h] 
\begin{center}
    \caption{ Comparative results of the proposed NADiffuSE on Auxialiary information. All results adopted the original interpolation operation where 20\% noisy signal is added at the end of the reverse process. Sign * means reproduced results.}
    \label{tab:condi}
    \begin{tabular}{c|c|cc|cccc}
    \hline
    \multirow{2}*{Row-Id}& \multirow{2}*{Method}& \multicolumn{2}{c|}{Auxiliary}& \multicolumn{4}{c}{Metrics}\\
    \cline{3-8}&&Spectrogram & Noise encoding & CSIG& CBAK & COVL & PESQ\\
    \hline
    0&Unprocessed&\usym{2717}&\usym{2717}&3.65&3.16&2.91&2.13\\
    \hline
    1&DiffuSE*&80-dim Mel+513-dim Spec&\usym{2717}&3.80&3.10&3.15&2.52\\
    \hline
    2&CDiffuSE*&80-dim Mel+513-dim Spec&\usym{2717}&3.83&3.13&3.19&2.55\\
    \hline 
     3&no-condition&\usym{2717}&\usym{2717}&2.87&2.61&2.36&1.88\\
     4&mel-conditioned&80-dim Mel+80-dim Mel&\usym{2717}&3.86&3.13&3.22&2.58\\
     5&noise-class&\usym{2717}&category classifier&3.52&2.86&2.85&2.20\\
     \hline
     6&NADiffuSE&80-dim Mel+80-dim Mel&hard classifier&\textbf{3.91}&\textbf{3.15}&\textbf{3.26}&\textbf{2.63}\\
     \hline 
     \end{tabular}
\end{center}
\vspace{-2em}
\end{table*}

\section{Experiment}
\subsection{Experimental Setup}
\subsubsection{Dataset}
\label{dataset}
We evaluated above all methods on the original and newly simulated VoiceBank-DEMAND \cite{c21} dataset. The new one follows the same setting which consists of 30 speakers from the VoiceBank \cite{c22} corpus and was mixed with 12 noise types\footnote{TCAR, PSTATION, DLIVING, TBUS, TMETRO, OMEETING, SPSQUARE, STRAFFIC, PRESTO, OOFFICE, PCAFETER, NFIELD.} from the DEMAND \cite{c23} database. It was then divided into a training, validation and test set with 26, 2 and 2 speakers, containing 10792, 770 and 824 synthesized utterances, respectively. The signal-to-Noise\,(SNR) range of the training and validation set is $\left\{0,5,10,15\right\}$, and the test set is $\left\{2.5,7.5,12.5,17.5\right\}$. All of the utterances were resampled to 16KHz sampling rates. We use PESQ, CSIG, CBAK and COVL as evaluation metrics for enhanced speech, with higher scores indicating better performance.\par
\subsubsection{Model infrastructure}
Our proposed model follows the generator-plus-conditioner architecture mentioned in the previous section. The generator is constructed based on DiffWave \cite{kong2020diffwave}, which is a diffusion model based vocoder and has been extended to the task of SE by \cite{diffuse,cdiffuse,modified}. The network in above works constructs from WaveNet \cite{oord2016wavenet} which has 30 layers of residual blocks with dilated convolution\,(conv) and gated activation. As shown in Fig. \ref{fig:overview} (d), each residual block has two 1x1 conv layers for Mel spectrograms and noise encodings.

\subsubsection{Training settings}
All of experiments are based on the same training configurations as CDiffuSE-base \cite{cdiffuse} in which 50-step linear noise scheduler $\beta_t\in [1\times10^{-4},0.035]$ is used. The learning rate is $2\times 10^{-4}$ and the batch size is 16 for both training stages. The dimension for the Mel spectrogram is 80, which is transformed by STFT with the window size of 1024 and the shift of 256. We uniformly train 100w iterations in the first training stage and 30w in the second stage. The full sampling scheme is used in the reverse process.

\subsection{Evaluation results for noise encodings}

\begin{figure}
\centering
\centerline{\includegraphics[width=\columnwidth]{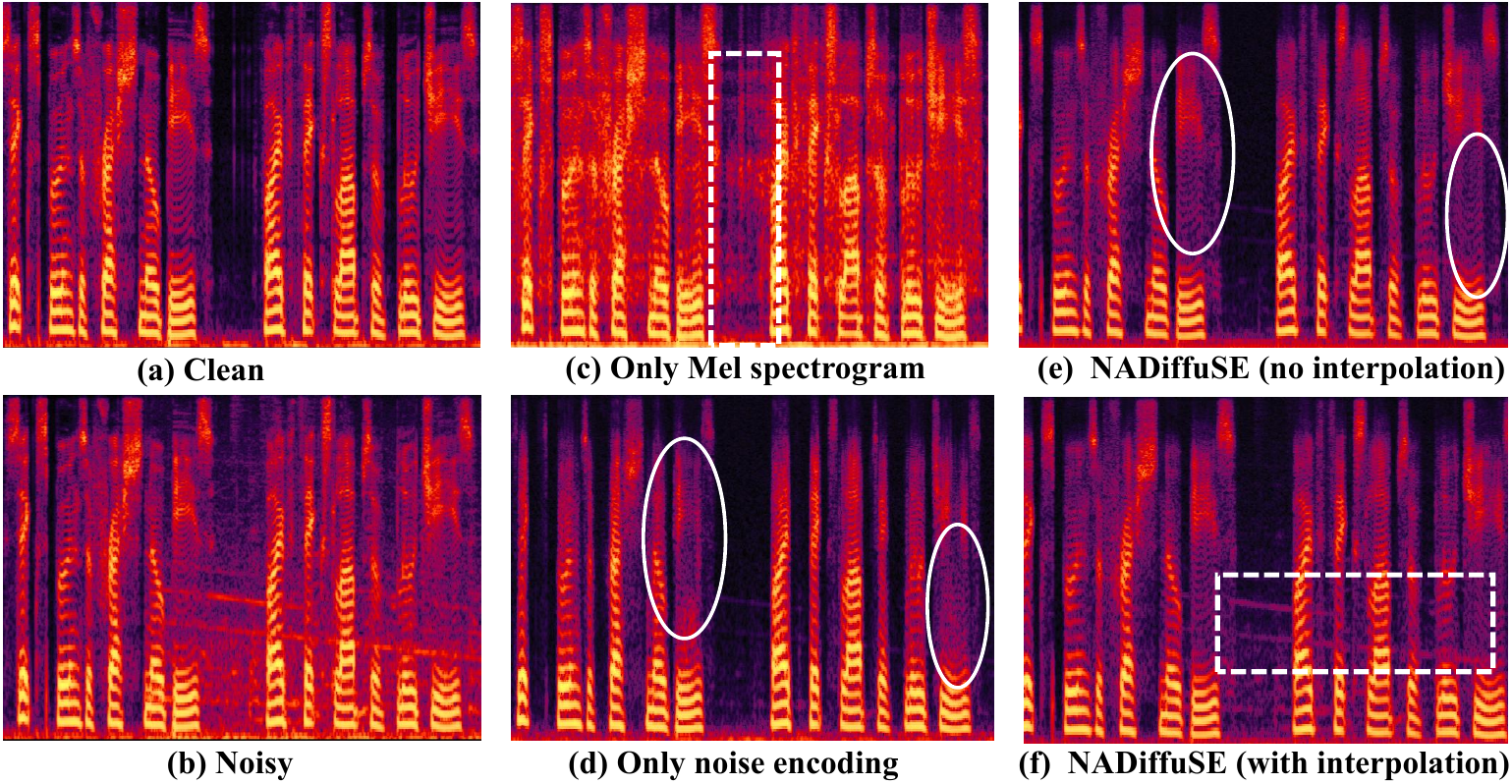}}
\caption{Mel spectrograms of (a) clean speech, (b) noisy speech, (c) enhanced speech only using the mel-spectrogram as condition, (d) enhanced speech only using noise encodings, (e) NADiffuSE's enhanced speech without the interpolation and (f) NADiffuSE's enhanced speech with the interpolation. The dashed box indicates noise residuals and the solid ellipse indicates speech distortion.}
\label{melset} 
\vspace{-1em}
\end{figure}

To resolve the non-Gaussian estimation problem proposed in the Section \ref{intro}, two types of noise encodings, namely category classifier\,(Cls) and latent embedding\,(Emb) were explored in Table \ref{t1}. Among them, the latent embedding form cannot obtain good performance gain in upper bound probably because it contains too much redundant information about audio events in the background noise that does not need to be considered. Therefore, our following proposed model uses the noise representation in the form of one hot vectors by default. In order to get the ground-truth background noise labels, we conducted experiments on the new simulated dataset. The ablation studies are done to validate the effectiveness of noise encodings. From Table \ref{tab:condi} we can observe: (1) Under the GPC archietecture, mel-spectrogram works as an important condition to improve the metric scores of restored speech signals when compared with row 3 and 4; (2) Metrics in row 5 increase a little compared to row 3, showing that the noise embeddings can improve the enhancement performance to some extent, but slightly lacks in speech detail fidelity; (3) The best performance is obtained when two conditions are both used as NADiffuSE\,(row 6) reach the highest scores among all combination of auxiliary information. It can be explained that the noise embedding provides the priori information about noise patterns, and thus viewed as an implicit multi-branch switch that guides the model for adaptive noise reduction. We also visualized the effects in in Fig. \ref{melset}: when inference only with Mel spectrograms, more details in the input noisy speech signal are reserved, but also including residual background noise components according to the dashed box in (c); when only using noise encodings, the noise is removed more completely, owing to the explicit use of noise characteristics, but some speech details are lost as shown by the solid ellipse in (d). Another point of interest is that the difference between Table \ref{tab:condi} row 2 and 4 is only in the spectral information used in the second training phase. The results show that row 4 can achieve comparable (or even a little better) results than row 2, which validates our view that there is no need to change the type and dimensionality of spectrograms before and after the two training phases, and lays the foundation for our proposal of a preprocessing network for the Mel spectrogram later on.

\subsection{Evaluation results for the improved inference}

\begin{figure}
\centerline{\includegraphics[width=\columnwidth]{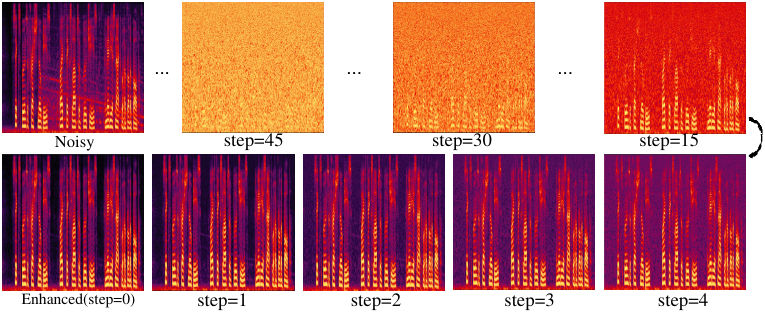}}
\caption{Visualization of the iterative reverse stage. The spectrograms of specific time steps are given, showing the importance of the last five steps for speech restoration.}
\label{fig:vis}
\end{figure}

\begin{table}[t]
\begin{center}
    {
    \caption{ Comparative results of the the anchor-based inference algorithm. For the sampling option, "$t=0$" with "$r=0.2$" means the original interpolation during inference\,(denoted by "-"). Sign "+" means our improved inference algorithm. All baseline are reproduced on the new dataset.}
    \scalebox{0.85}{
    \label{tab:inf}
    \begin{tabular}{c|c|cc|cccc}
    \hline
    \multirow{2}*{Row-Id}& \multirow{2}*{Method}&\multicolumn{2}{c|}{Sampling}& \multicolumn{4}{c}{Metrics}\\
    \cline{3-8}&& Ratio (r)& Step (t) &CSIG & CBAK & COVL & PESQ\\
    \hline
    0&Unprocessed&\usym{2717}&\usym{2717}&3.65&3.16&2.91&2.13\\
    1&SEGAN&\usym{2717}&\usym{2717}&3.71&3.12&3.04&2.35\\
    2&SGMSE+&\usym{2717}&\usym{2717}&\textbf{4.18}&\textbf{3.47}&\textbf{3.58}&\textbf{3.24}\\
    3&DiffuSE&r=0.2&t=0&3.80&3.10&3.15&2.52\\
    4&CDiffuSE&r=0.2&t=0&3.83&3.13&3.19&2.55\\
    5&CDiffuSE-&\usym{2717}&\usym{2717}&3.80&3.13&3.16&2.52\\
    6&NADiffuSE&r=0.2&t=0&3.91&3.15&3.26&2.63\\
    7&NADiffuSE-&\usym{2717}&\usym{2717}&3.84&3.09&3.18&2.55\\
     \hline 
    \multirow{3}*{8}&\multirow{3}*{NADiffuSE+}&r=0.2&\multirow{3}*{t=5}&3.92&3.15&3.25&2.62\\
     &&r=0.5&&3.81&3.11&3.09&2.37\\
     &&r=0.8&&3.77&3.15&2.92&2.15\\
     \hline
     9&NADiffuSE+&r=0.1&t=5&\textbf{3.94}&\textbf{3.18}&\textbf{3.31}&\textbf{2.69}\\
     \hline
     \multirow{3}*{10}&\multirow{3}*{NADiffuSE+}&\multirow{3}*{r=0.1}&t=50&3.88&3.12&3.25&2.63\\
     &&&t=10&3.89&3.13&3.26&2.64\\
     &&&t=2&3.91&3.16&3.27&2.64\\
     \hline
     \end{tabular}
     }
     }
\end{center}
\vspace{-2em}
\end{table}
\begin{table}[t]
\begin{center}
    \caption{ Evaluation Results of NADiffuSE+ and its variants. For the needed noise labels, we need an accurate noise classifier. We denote the ground-truth noise label as gt. Models with the 92\% and 96\% accuracy classifier are abbreviated as 92\%-NADiffuSE+ and 96\%-NADiffuSE+.}
    \label{tab:variants}
    \begin{tabular}{cccccc}
      \hline
        Row-Id &Method  & CSIG & CBAK & COVL & PESQ\\
        \hline
        0&\textit{gt-NADiffuSE+}&3.95&3.22&3.31&2.69\\
        1&\textit{92\%-NADiffuSE+}&3.94&3.18&3.29&2.66\\
        2&\textit{96\%-NADiffuSE+}&3.94&3.18&3.31&2.69\\
        3&\textit{Coarse-and-refine}&\textbf{4.06}&\textbf{3.22}&\textbf{3.44}&\textbf{2.84}\\
        4&\textit{Coarse-and-finetune}&3.93&3.20&3.24&2.54\\
        5&\textit{Coarse-and-scratch}&3.98&3.19&3.35&2.74\\
      \hline
     \end{tabular}
\end{center}
\vspace{-2em}
\end{table}
To evaluate the anchor-based inference algorithm, we first analyze the original interpolation operation in the inference algorithm and further investigate different parameter settings for the improved inference algorithm. We can first summarize from Table \ref{tab:inf} that without any interpolation operation, the performance will significantly degrade because there is a big gap between NADiffuSE- and NADiffuSE. As we have analyzed in Section \ref{method:inf}, we only use iterative interpolation operations in a limited number of time steps because of the increasing sensitivity to extra noise during the inference. In Fig. \ref{fig:vis}, we give an example of visualization on the inference process where the last five steps play an important role in restoring the speech signal. Therefore, we empirically choose $t=5$ first and explore the effect of different interpolation coefficients on the final performance in the row 8, and we can find that the smaller the coefficient is, the higher the performance gain can be obtained. Specially, we degrade the coefficients linearly on average with the time step, e.g. for $t=5,r=0.1$, we would apply coefficients of $[0.1,0.08,0.06,0.04,0.02]$ for the last five steps respectively. After we experimentally selected r = 0.1, we did a validation test in row 10 and the results showed that the final performance indeed decreases as the time step t increases. Ovarally, we have some conclusions: (1) Our proposed inference algorithm is more effective than the original one. (2) $r=0.1,t=5$ is the best parameter setting. 

\subsection{Evaluation results for the proposed model and its variants}

\begin{table}[t]
\begin{center}
    \caption{ Results of unseen noise scenes from the simulated dataset with VoiceBank and Noisex-92.}
    \label{noisex-92}
    \begin{tabular}{ccccc}
    \hline
    Method  & CSIG & CBAK & COVL & PESQ\\
    \hline
     \textit{Unprocessed}&2.84&2.75&2.21&1.59\\
     \hline
     \textit{DiffuSE}&2.92&2.69&2.40&1.89\\
     \textit{CDiffuSE}&2.97&2.72&2.42&1.91\\
     \hline
     \textit{NADiffuSE+}&\textbf{3.04}&\textbf{2.73}&\textbf{2.46}&\textbf{1.92}\\
      \hline
     \end{tabular}
\end{center}
\vspace{-2em}
\end{table}
\subsubsection{Overal model}
Through the previous experiments, our proposed model is identified as a time-domain diffusion model that uses Mel specrograms and the category classifier form of noise encodings as conditional information in both training stages. For inference, we need to train a noise classifier\,(detail in Section \ref{method:classifier}) to get the estimated noise type from the given noisy speech signal. The experimental results in Table \ref{tab:variants} show that the convergence of the noisy classifier on our newly simulated dataset is not difficult and there is no significant difference in the final SE performance between the two trained classifiers with 92\% and 96\% accuracy\,(row 1 v.s. 2). This can be interpreted as redundancy in the current category labeling of background noise, which inspires us to explore more fine-grained noise encoding. Our proposed NADiffuSE and NADiffuSE+ both perform better than DiffuSE \cite{diffuse} and CDiffuSE \cite{cdiffuse}. But this way of diffusion-based SE models still lag behind spectral domain methods such as \cite{sgmse+}, which uses the stochastic differential equation to formulate the diffusion process. Furthermore, we also conduct out-of-domain experiments to validate NADiffuSE's generalization ability for unseen noise scenes compared with baselines. We simulated the new dataset using VoiceBank \cite{c22} and Noisex-92 with the same setting in Section \ref{dataset}. Results in Table \ref{noisex-92} show that our method achieves better scores than other time-domain models under the GPC structure. 

\subsubsection{Variants}
We compare the performance of the three network variants described in the previous paper and all our inferences are based on the improved algorithm with the best parameter setting, $r=0.1,t=5$. All variants are based on 96\%-NADiffuSE+. As shown in Table \ref{tab:variants}, the coarse-and-refine approach achieves the best performance. The coarse-and-scratch can get a small performance gain with only one training stage required. The coarse-and-finetune performs comparably to the originally proposed model, with no significant performance gains, which may further need a well-selected pre-trainig checkpoint and fine-tuning strategy. We can gain insight from row 3 that multi-stage iterative speech enhancement can help recover speech details better, and generative diffusion models have great potential for detail refinement in data reconstruction. All above network variants essentially rely on the training of respective sub-modules, so the quality of the pre-enhanced spectrogram feature would limit the final SE performance.

\section{Conclusions}
We summarize and analyze the current diffusion-based speech enhancement methods, where in the setting of generator-plus-conditional architecture\,(GPC), we propose a noise-aware diffusion-based SE model\,(NADiffuSE), which conducts denoising under the global guidance of noise encodings to help the non-Gaussian noise estimation. To reduce the additional noise introduced by the original interpolation operation, we propose the anchor-based inference algorithm to to complement speech details and reduce the residual noise. Plus, we investigate three variants of NADiffuSE which use the preprocessing network to enhance the Mel spectrogram in advance, to further bridge the gap in the performance bounds. Through experiments, we have shown that our model performs better than other diffusion-based SE models under the GPC structure.

\section*{Acknowledgment}
This paper was partially supported by NSFC (No:62176008) and Shenzhen Science\&Technology Research Program (No: GXWD20201231165807007-20200814115301001).


\end{document}